\documentclass[aps,prl,twocolumn,amsmath,amssymb, superscriptaddress,tightenlines]{revtex4}
\usepackage{graphicx}% Include figure files
\usepackage{epsfig}
\usepackage{dcolumn}% Align table columns on decimal point
\usepackage{bm}% bold math
\usepackage{amssymb}
\usepackage{bm}
\usepackage{mathtools}
\usepackage{amsmath, bm}
\usepackage{upgreek}
\usepackage[colorlinks,citecolor=blue,linkcolor=blue,hyperindex]{hyperref}

\begin{document}

\title{Kolmogorov complexity as intrinsic entropy of a pure state: Perspective from entanglement in free fermion systems}
\author{Ken K. W. Ma and Kun Yang}
\affiliation{National High Magnetic Field Laboratory and Department of Physics, Florida State University, Tallahassee, Florida 32306, USA}
\date{\today}

%----------------------- Abstract ------------------------------

\begin{abstract}
We consider free fermion systems in arbitrary dimensions and represent the occupation pattern of each eigenstate as a classical binary string. We find that the Kolmogorov complexity of the string correctly captures the scaling behavior of its entanglement entropy (EE). In particular, the logarithmically-enhanced area law for EE in the ground state and the volume law for EE in typical highly excited states are reproduced. Since our approach does not require bipartitioning the system, it allows us to distinguish typical and atypical eigenstates directly by their intrinsic complexity. We reveal that the fraction of atypical eigenstates which do not thermalize in the free fermion system vanishes exponentially in the thermodynamic limit. Our results illustrate explicitly the connection between complexity and EE of individual pure states in quantum systems.
\end{abstract}

\maketitle

%-------------------------- Introduction -----------------------------

\section{Introduction} 

Statistical mechanics of isolated quantum system is a topic of tremendous current interest~\cite{Deutsch, Srednicki, Tasaki, Olshanii, Polkovnikov-RMP, Rigol-review, Deutsch2013, Gogolin-thesis, Eisert1, Eisert2, Huse-review, Grover-PRX, Schmiedmayer}. Without an external heat bath, definitions of standard thermodynamic quantities such as temperature and entropy become subtle and often ambiguous in such systems~\cite{Deutsch2013, Goold, Haque-Temp}. For example, every pure state has zero von Neumann entropy~\cite{vN}. On the other hand, according to the eigenstate thermalization hypothesis (ETH)~\cite{Deutsch, Srednicki, Tasaki, Olshanii}, a highly excited eigenstate of a nonintegrable system should be ``thermal" and thus have the same temperature and corresponding entropy density of a (mixed) thermal state with the same energy density (and other conserved charge densities if present). One way to resolve this tension is to partition the system (usually in position space) and focus on the smaller subsystem, which is in a mixed state and has a nonzero von Neumann entropy associated with its reduced density matrix. This is known as the entanglement entropy (EE). Indeed, it has been shown that the overwhelming majority of (or typical) free fermion eigenstates give rise to thermal reduced density matrices, a property termed eigenstate typicality~\cite{Yang2015,Yang2018}. An immediate consequence is that EE equals the corresponding thermal entropy in these cases. Eigenstate typicality also plays an important role in the dynamical generation of entanglement in free fermion systems~\cite{TianYang}.

This (by now standard) way of revealing the thermal nature of a pure state is unsatisfactory in several aspects. First of all, in principle EE depends on the way the system is partitioned, while entropy should be an {\em intrinsic} property of a state. Free fermion states are good examples of this: They are highly entangled in real space, but are product states with zero EE in a momentum space partitioning. Second, entanglement is a unique property of quantum mechanics~\cite{Horodecki-RMP}, while the notion of entropy was first introduced in classical statistical thermodynamics, where all {\em  individual} (or micro; {\em not} an ensemble of) states are pure. As a result, the von Neumann definition of entropy would {\em always} be zero there, regardless of whether one considers the whole universe or a subset of it. While one may object that the universe is intrinsically quantum, we can always consider semiclassical pure states that are well-described by classical physics, whose EE can be made arbitrarily small.

Over the years, various alternative definitions of entropy have been introduced, in an attempt to reveal the {\em intrinsic} thermal properties of a state, either mixed or pure~\cite{Deutsch2010, Baez, Deutsch-en, Winter-PRXQ, Holecek, Zurek-nature, Zurek1989, Zurek-book, Caves}. Meanwhile, the applications of classical and quantum Kolmogorov complexity make it possible to quantify the complexity of quantum states~\cite{Svozil, Vitanyi2000, Vitanyi2001, Gacs, Berthiaume, Yamakami, Benatti, Benatti2, Mora2006, Mora2007, Mora-PRL, Mueller-thesis, Rogers, Eisert2016, Susskind1, Susskind2, Aguero, Eisert2021, Hnilo, Bhojraj, Kazemi, Kaltchenko, Mueller-QKC-KC}. It has been suggested that physical entropy should be a reflection of the complexity of a state, and quantify the amount of information carried by (or ``hidden" in) it~\cite{Zurek-nature, Zurek1989, Zurek-book, Caves, Susskind1, Susskind2}. Furthermore, it has been shown that the von Neumann entropy of a probabilistic source (or density matrix) and the average quantum Kolmogorov complexity of the qubit strings generated by the source should coincide~\cite{Svozil, Vitanyi2000, Vitanyi2001, Gacs, Berthiaume, Yamakami, Mueller-thesis, Benatti, Benatti2, Rogers}. Nevertheless, a concrete example of the connection between Kolmogorov complexity and nonzero EE of individual pure states remains elusive.

In this paper, we show that the \textit{classical} Kolmogorov complexity of free fermion states has the same scaling behavior as their bipartite EE, thus directly relating EE to the intrinsic complexity of such pure states. Furthermore, Kolmogorov complexity is a quantitative measure of how typical a state is. This not only provides a systematic way to distinguish between typical and atypical eigenstates in the free fermion system from their occupation patterns, but also allows us to demonstrate that the fraction of atypical eigenstates which do not thermalize in the free fermion system vanishes exponentially in the system size in the thermodynamic limit. Our results shed light on the quantification of typical and atypical (the non-interacting version of scar~\cite{Serbyn-review}) states, which is important in understanding thermalization and the emergence of statistical mechanics in pure states. 

\section{A brief review of Kolmogorov complexity} 

Given a binary string $x$, its Kolmogorov complexity is defined as the length of the shortest possible description of $x$~\cite{Solomonoff1964, Kolmogorov1965, Chatin1987, Li-Vitanyi}. Specifically, one can consider the ``two-part codes" which consist of a universal Turing machine and a program~\cite{Li-Vitanyi}. Then, the \textit{plain} Kolmogorov complexity of $x$ is defined as~\cite{footnote-K}
\begin{eqnarray}
C(x)=\text{min}\left\{l(T)+l(p):T(p)=x
\right\}+O(1).
\end{eqnarray}
The program $p$ is executed by the universal Turing machine $T$, which outputs the string
$x$ and halts. Here, $l(p)$ denotes the length of $p$ in bits. It is obvious that the shortest possible program that can reconstruct $x$ depends on the choice of $T$. Nevertheless, using another Turing machine (or computer) can only lead to a difference in $C(x)$ bounded from above by a finite constant that is independent of $l(p)$. In other words, this is a change in $O(1)$. Furthermore, the length of the self-delimiting encoding of $T$, i.e. $l(T)$, is independent of $l(p)$. Therefore, it is common to simply focus on $l(p)$ and view it as the Kolmogorov complexity of $x$. Roughly speaking, all irregularities in the string $x$ are reflected by $l(p)$.

With the above definition, we now summarize some important results for $C(x)$. Although the value of $C(x)$ cannot be computed from any program, $C(x)$ is bounded from above. Consider a string which is random and has no simple description. To output the string, the best one can do is to take the entire string as the input and ask the Turing machine to copy the input to the output. Hence, the Kolmogorov complexity of any string satisfies
$C(x)\leq l(x)+O(1)$. It is expected that a typical string is random and has $C(x)\simeq l(x)$. We use the symbol $\simeq$ when the relationship holds up to the leading order. As the $O(1)$ term becomes negligible for sufficiently long strings, it will be dropped for convenience.

On the other hand, some strings are easy to describe. For example, consider the string $``11\cdots 1"$ where the bit $``1"$ is repeated $n$ times. We abbreviate the string as $1^n$. This abbreviation immediately shows that the string is very simple and can be reconstructed from a very short input. Specifically, one can define a Turing machine which prints $``1"$ for $n$ times. Now, we simply need $\log{n}$ bits to specify the binary representation of $n$ in the program $p$~\cite{footnote-log}. Alternatively, we can say that the string $1^n$ is highly compressible by encoding it as the binary representation of $n$. Hence, the string $1^n$ has Kolmogorov complexity,
\begin{eqnarray} \label{eq:C-log}
C(1^n)\simeq \log{n}.
\end{eqnarray}

A string $x$ is called $c$-incompressible if its Kolmogorov complexity satisfies $C(x)\geq l(x)-c$. Note that the upper bound $C(x)\leq l(x)+O(1)$ always holds. Denote the set of all binary strings as $\mathcal{B}=\left\{\Lambda, 0, 1, 00, 01, 10, 11, \cdots\right\}$, where $\Lambda$ is the empty string. The total number of binary strings with lengths shorter than $N-c$ is
\begin{eqnarray} \label{eq:c-incompressible}
\sum_{i=0}^{N-c-1} 2^i
=2^{N-c}-1.
\end{eqnarray}
When one encodes $x$, the final result must be an element in $\mathcal{B}$. Notice that different elements in $\mathcal{B}$ may correspond to different encodings of the same string. Hence, the largest possible fraction of strings with length $N$ that is $c$ compressible is
\begin{eqnarray}
\frac{2^{N-c}-1}{2^N}=2^{-c}~,~ \text{for } N\rightarrow\infty.
\end{eqnarray}
This result implies that most of the strings cannot be compressed by a significant amount. Therefore, simple strings do exist but they are rare and atypical.  Furthermore, $x$ is said to be Kolmogorov random if it cannot be compressed by one bit. From the pigeonhole principle~\cite{pigeonhole}, there must be at least one string for every length $N$ that is incompressible.

Moreover, the difficulty in describing $x$ depends on the information $y$ that is already specified to the program. This leads to the concept of conditional Kolmogorov complexity, denoted as $C(x|y)$. The difference between $C(x)$ and $C(x|y)$ is the most noticeable in simple strings. For example, suppose that the length of the string $N$ is given. Then, $1^{N}$ has a conditional Kolmogorov complexity,
\begin{eqnarray} \label{eq:C0}
C(1^N| N)= c,
\end{eqnarray}
where $c$ is a constant. Another related example for our later discussion is the string which has a fixed number of ones in its elements. When both the length of the string $N$ and the number of ones in the string $n$ are given, then $C(x|N,n)\lesssim\log{\binom{N}{n}}$.
Applying Stirling's approximation, one has
\begin{eqnarray} \label{eq:K_alpha}
C(x|N,n)\lesssim N H(n/N).
\end{eqnarray}
Here,
\begin{eqnarray}
H(\alpha)=-\alpha\log\alpha-(1-\alpha)\log(1-\alpha)
\end{eqnarray}
is the Shannon entropy of a Bernoulli distribution~\cite{Shannon}. 

\section{Entanglement entropy in a 1D free fermion system} 

In gapped systems described by local Hamiltonians and most of the gapless systems in $d>1$ dimensions, ground state EE satisfies an area law and scales with the surface area of the subsystem, $S\sim L^{d-1}$~\cite{Hastings, Amico-RMP, Eisert-RMP}. This originates from local or short-distance entanglement. Here, $d$ is the dimensionality of the system and $L$ is the typical length of the subsystem in any direction. However, EE of free fermions in the ground state satisfies $S\sim d L^{d-1}\log{L}$~\cite{GK2006, Wolf2006, Swingle2010, Ding-PRX, footnote-violation}, whereas a volume law $S\sim L^d$ is satisfied in the vast majority of highly excited eigenstates~\cite{Yang2015}.

We first consider the system of $n$ free spinless fermions in one dimension, and show that the Kolmogorov complexity of eigenstates has the same scaling behavior as their bipartite EE. We assume that there are $N$ different single-particle eigenstates in the momentum space, where $N$ is proportional to the volume (in $1$D, length) of the system. In the following discussion, we are only interested in the thermodynamic limit, in which both $n$ and $N$ are infinite but the ratio $\alpha=n/N$ is fixed.

Now, each many-body eigenstate can be described by an occupation pattern $(n_1, n_2, \cdots, n_N)$. Here, $n_i=1$ if the single-particle eigenstate with momentum $k_i$ is occupied by a fermion. Otherwise, $n_i=0$. This description resembles a binary string $x$ with length $N$ that has $n$ ones in its elements. From Eq.~\eqref{eq:K_alpha}, the Kolmogorov complexity of a typical occupation pattern is asymptotically equal to the Shannon entropy, i.e. $C(x|\alpha)\simeq NH(n/N)$. The scaling behavior of $C(x|\alpha)$ agrees with the volume law of EE in typical eigenstates~\cite{Yang2015}. Here, we reemphasize that EE in these states is also the thermal entropy since the typical eigenstates are thermal~\cite{Yang2015}. Later, we will have further discussion of typical and atypical eigenstates.

What happens if we apply the above argument to the ground state of the system? Suppose that the single-particle eigenstates with the $n$ smallest momenta are occupied. This occupation pattern leads to the binary string $1^n 0^{N-n}$. From Eq.~\eqref{eq:C-log}, this string has a Kolmogorov complexity
$C(x_\text{GS}|\alpha)\simeq \log{n} \simeq \log{N}\propto \log{L}$. Here, $x_{\text{GS}}$ denotes the ground state occupation pattern. For a more generic Hamiltonian, the free-fermion ground state may possess $m>1$ Fermi surfaces (pairs of points in 1D). When $m\ll n$, the Kolmogorov complexity of the occupation pattern satisfies $C(x_\text{GS}|\alpha)\simeq m\log{N}$~\cite{footnote-many}. The above results correctly reproduce the scaling behavior of the ground state EE without bipartitioning the system! Since the thermal entropy should be extensive and scale as $L$ in one dimension, the EE of the ground state is not the thermal entropy. In both ground state and typical eigenstates, the Kolmogorov complexity of the occupation pattern agrees with the scaling behaviors of EE. 

\section{Entanglement entropy in higher dimensional free fermion systems} 

\begin{figure} [htb]
\begin{center}
\includegraphics[width=2.5in]{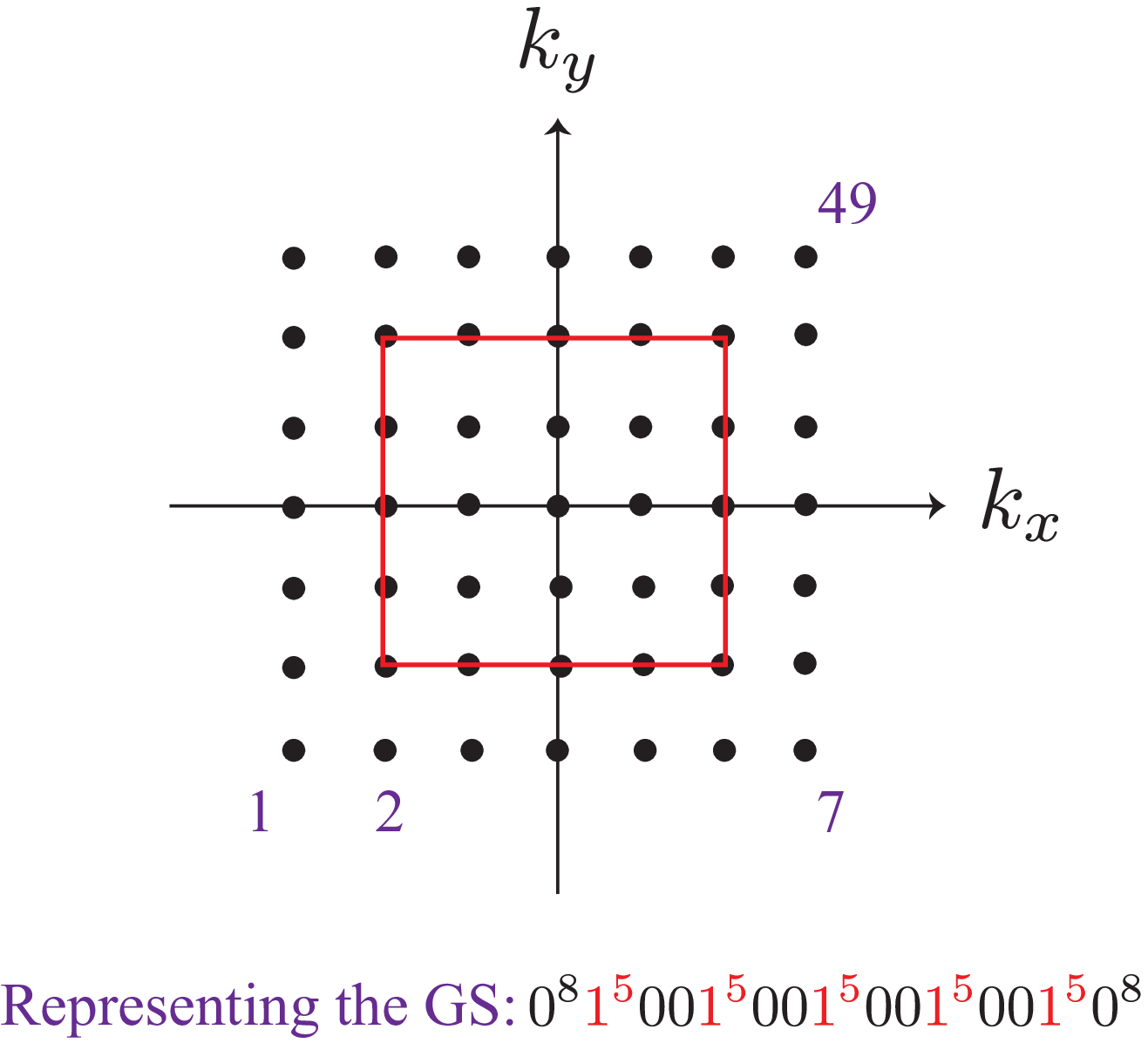}
\caption{Labeling single-particle eigenstates (the black dots) and representing the ground state (GS) occupation pattern in the two-dimensional free fermion system as a binary string. Here, the square Fermi surface is a subgraph of the complete graph formed by the $49$ vertices. All the edges in the complete graph are skipped for better illustration. To describe the Fermi surface, one needs to specify the labels $9-13$, $16-20$, $23-27$, $30-34$, and $37-41$ in order of their connection.}
\label{fig:FS-string}
\end{center}
\end{figure}

For free fermion systems in $d>1$ dimensions, we can still assign a label to each single-particle eigenstate in the momentum space. Each label takes a value between $1$ to $N$, with the values of all labels being different. Fig.~\ref{fig:FS-string} illustrates an example of labeling the single-particle eigenstates in the two-dimensional momentum space. We assume that the labeling scheme is a piece of information that is already specified to the program. Now, it becomes very straightforward to generalize the previous discussion on typical eigenstates to $d>1$ dimensions. Again, the occupation pattern for a typical eigenstate satisfies
$C(x|\alpha)\simeq NH(\alpha) \propto L^d$. Here, $L^d$ is the volume of the system. Just as with the one-dimensional system, the result resembles the volume law of EE in a typical eigenstate.

One may naively think that the previous argument on the ground state in one-dimensional system can also be directly generalized to higher dimensions. This will lead to a Kolmogorov complexity that scales as $\log{N}\propto d\log L$, which does not agree with the EE of the ground state, $S\sim dL^{d-1}\log{L}$~\cite{GK2006, Wolf2006, Swingle2010, Ding-PRX}. However, the naive generalization breaks down because the occupation pattern $1^n0^{N-n}$ contains no information about the shape of the Fermi surface (FS)! Therefore, we need to develop a suitable approach for describing the FS with its shape in the form of a one-dimensional binary string.

The tool we employ is graph theory~\cite{Diestel-graph}. A graph is an ordered pair $G(V,E)$ comprising a set of vertices $V$ and a set of edges $E$. A graph $G'(V',E')$ is a subgraph of $G$ if and only if $V'\subseteq V$ and $E'\subseteq E$. Usually, one represents the adjacency matrix as a binary string to describe the graph. Since there are at most $N(N-1)/2$ edges, the Kolmogorov complexity of a typical graph scales as $N^2$. Meanwhile, basic (simple) graphs exist. An example is the complete graph, in which each vertex is connected to all other vertices. This graph has Kolmogorov complexity $O(1)$~\cite{K-graph}. Now, the FS encloses $n$ points (including the points on the FS) in the momentum space. In particular, the points on the FS and the edges connecting them form a cycle subgraph of the aforementioned complete graph. Obviously, this subgraph describes the shape of the FS. To describe this cycle subgraph, we need to specify the labels of the vertices on the FS in the order of their connection~\cite{K-graph}. It takes no more than $\log{N}\propto\log{L^d}$ bits to specify each label. An example for the two-dimensional system is given in Fig.~\ref{fig:FS-string}. We assume that the system is nearly isotropic, such that the number of vertices lying on the FS scales as $n^{(d-1)/d}\sim L^{d-1}$. After describing the FS, a suitably defined Turing machine can fill in $1$ for the string elements which label the points inside the FS and $0$ otherwise. Therefore, the occupation pattern for the ground state has Kolmogorov complexity,
\begin{eqnarray} \label{eq:K-gs}
C(x_\text{GS}|\alpha)\simeq n^{(d-1)/d}\log N
\propto d L^{d-1}\log L.
\end{eqnarray}
This agrees with the scaling behavior of EE of free fermions in the ground state~\cite{GK2006, Wolf2006, Swingle2010, Ding-PRX}.

What happens if $n=N$, corresponding to a band insulator? In this case, all single-particle eigenstates are occupied, and there is no FS. From Eq.~\eqref{eq:C0}, we know that the occupation pattern has $C(x|n)= O(1)$. The occupation pattern can be specified by describing the cycle graph that connects all outermost vertices, which has $C(G)=O(1)$~\cite{K-graph}, consistent with the simple result above. In this case, EE is actually dominated by local or short-distance entanglement (not directly related to the complexity of the global state) that gives rise to the area law.

It is illuminating to compare the above case with disordered free fermions, where there is no FS even for the metallic phase. In this case the ground state EE always satisfies the area law~\cite{Potter, Pouranvari}.  Since momentum is no longer a good quantum number, the previous graph theoretic description of the FS becomes unsuitable. Instead, each single-particle eigenstate is labeled by its eigenenergy. The Kolmogorov complexity of the global many-body ground state scales as $\log{N}\sim d\log{L}$ just as in the 1D case, but is subdominant compared with the area law contribution. We conjecture that this logarithmic term would show up as a subleading correction in the EE in the metallic phase, while it is absent in the insulating phase. It would be very interesting to  test this numerically.

Although our focus in this paper is the free fermion system, the methodology can be easily generalized to other systems. An obvious example is free boson states. At zero temperature all bosons condense into momentum $\mathbf{k}=\mathbf{k}_0$~\cite{footnote-generic}. This ground state can be described by specifying the label of the single-particle eigenstate (see Fig.~\ref{fig:FS-string}) being occupied by the bosons. Thus, $C(x_{\text{GS}}|\alpha)\simeq\log{N}$. This result agrees with the scaling behavior of bipartite EE obtained in Refs.~\cite{Klich2006, Ding2009}, and there is no area-law contribution in this case. Suppose that the system is perturbed by a weak interaction between bosons. In this (more generic) case there is an area law term in the ground state EE, while the logarithmic term from the condensate becomes a subleading contribution, which comes from the spontaneously broken continuous symmetry it represents and the corresponding quantum fluctuations of the order parameter and Goldstone modes~\cite{Ding2008, Grover}. Such behavior is consistent with the scenario described in the paragraph above.

It is worthwhile to mention that such subleading contributions are in some sense more important than the leading area law contribution in EE, as they reflect the {\em intrinsic} complexity of the global state. A famous example is the topological EE~\cite{Wen-TEE, Kitaev-TEE}, which captures the topological nature of the ground state. 

\section{Typical and atypical eigenstates} 

Previously, we observed that it is much easier to describe the occupation pattern for the ground state than the typical eigenstates for free fermions. Now, we make the distinction more explicit. We define a state as typical if and only if the Kolmogorov complexity of its occupation pattern scales as the number of particles or the system size. On the other hand, the occupation pattern of an atypical state has a Kolmogorov complexity that scales as $o(N)$. Since satisfying the volume law in EE is a necessary condition for thermalization, the above definition and our previous results directly imply that atypical states do not thermalize.

What can we say about the population of atypical eigenstates? Following the reasoning in Eq.~\eqref{eq:c-incompressible}, we know that the largest possible number of atypical eigenstates is $2^{o(N)+1}-1$. The number is actually smaller as the entanglement entropy of an eigenstate cannot be lower than that of the ground state. For the entire energy spectrum, there are $\binom{N}{n}\simeq 2^{NH(n/N)}$ different many-body eigenstates for the free fermion systems. Suppose that $\alpha=n/N$ is sufficiently away from $0$ or $1$, so that $H(n/N)$ is not close to zero. In the thermodynamic limit, the largest possible fraction of atypical eigenstates in the entire spectrum,
\begin{eqnarray} \label{eq:ratio}
\lim_{N\rightarrow\infty}
\frac{2^{o(N)+1}-1}{2^{NH(n/N)}}
\rightarrow 0,
\end{eqnarray}
vanishes exponentially. This justifies our definitions of typical and atypical eigenstates based on the Kolmogorov complexity of their occupation patterns.

\section{Conclusion and discussion} 

To conclude, using free fermion systems as our primary examples, we have demonstrated explicitly the connection between the intrinsic complexity and entanglement entropy of individual pure states. Specifically, we have shown that the Kolmogorov complexity of the fermion occupation pattern successfully reproduces the logarithmically-enhanced area law and the volume law of EE for the ground state and typical eigenstates, respectively. In the latter case, the Kolmogorov complexity asymptotically agrees with the Shannon entropy in the thermodynamic limit~\cite{Grunwald}.

Interestingly, our result suggests an alternative explanation to the logarithmic enhancement in the ground state EE. By representing the Fermi surface as a graph, the logarithmic term originates from the number of bits required to specify a vertex on the FS.

Furthermore, we distinguish between typical and atypical eigenstates by the Kolmogorov complexity of their occupation patterns. Based on this, we deduced that the fraction of atypical eigenstates in the entire spectrum vanishes exponentially in the thermodynamic limit. As pointed out in Ref.~\cite{Yang2018}, these atypical states can be easily eliminated by mixing with the typical states when the fermions interact. It is expected that most of the states in the interacting system would satisfy the volume law of EE and become thermal. On the other hand, quantum states with low EE, which are analogous to scar states~\cite{Serbyn-review}, may still persist with low probabilities. Our present approach cannot prove or disprove the strong ETH, which postulates that \textit{all} highly excited states in nonintegrable systems are thermal~\cite{strong-ETH}. We leave this important problem for future studies. 

Last but not least, we should clarify that the intrinsic complexity of a generic pure state may not be quantified by the classical Kolmogorov complexity. To serve the purpose, the concept of quantum Kolmogorov complexity was introduced~\cite{Li-Vitanyi, Svozil, Vitanyi2000, Vitanyi2001, Gacs, Berthiaume, Yamakami, Benatti, Benatti2, Mueller-thesis, Rogers, Mora-PRL, Mora2006, Mora2007}. Nevertheless, explicit examples of the connection between intrinsic complexity and entanglement entropy in realistic physical systems remain elusive. The simple free fermion system allows us to define its occupation patterns in momentum space which take a disentangled form (i.e., behave as classical-like objects), and quantify the intrinsic complexity of its eigenstates by classical Kolmogorov complexity. This further allows us to demonstrate its connection to the entanglement entropy. In fact, Kolmogorov complexity was employed in studying the physical entropy of classical systems, in particular, the Boltzmann gas~\cite{Zurek1989}. Its relevance to entanglement entropy in quantum systems that may have classical-like descriptions of their wave functions in some basis is revealed in this paper. Therefore, we believe that our work provides an important step in the research direction of connecting intrinsic complexity and entanglement entropy in (quantum) pure states. \\

\begin{acknowledgments}
This research was supported by National Science Foundation Grant No. DMR-1932796, and performed at the National High Magnetic Field Laboratory, which is supported by National Science Foundation Cooperative Agreement No. DMR-1644779, and the State of Florida.
\end{acknowledgments}


\begin{thebibliography}{99}

\bibitem{Deutsch}
J. M. Deutsch,
\href{https://journals.aps.org/pra/abstract/10.1103/PhysRevA.43.2046}
{Phys. Rev. A \textbf{43}, 2046 (1991)}.

\bibitem{Srednicki}
M. Srednicki,
\href{https://journals.aps.org/pre/abstract/10.1103/PhysRevE.50.888}
{Phys. Rev. E \textbf{50}, 888 (1994)}.

\bibitem{Tasaki}
H. Tasaki,
\href{https://journals.aps.org/prl/abstract/10.1103/PhysRevLett.80.1373}
{Phys. Rev. Lett. \textbf{80}, 1373 (1998)}.

\bibitem{Olshanii}
M. Rigol, V. Dunjko, and M. Olshanii,
\href{https://www.nature.com/articles/nature06838}
{Nature \textbf{452}, 854 (2008)}.

\bibitem{Gogolin-thesis}
C. Gogolin,
\href{https://arxiv.org/abs/1003.5058}
{arXiv:1003.5058}

\bibitem{Polkovnikov-RMP}
A. Polkovnikov, K. Sengupta, A. Silva, and M. Vengalattore,
\href{https://journals.aps.org/rmp/abstract/10.1103/RevModPhys.83.863}
{Rev. Mod. Phys. \textbf{83}, 863 (2011)}.

\bibitem{Rigol-review}
L. D'Alessio, Y. Kafri, A. Polkovnikov, and M. Rigol,
\href{https://www.tandfonline.com/doi/full/10.1080/00018732.2016.1198134}
{Adv. Phys. \textbf{65}, 239 (2016)}.

\bibitem{Eisert1}
C. Gogolin and J. Eisert,
\href{https://iopscience.iop.org/article/10.1088/0034-4885/79/5/056001}
{Rep. Prog. Phys. \textbf{79}, 056001 (2016)}.

\bibitem{Eisert2}
J. Eisert, M. Friesdorf, and C. Gogolin,
\href{https://www.nature.com/articles/nphys3215}
{Nat. Phys. \textbf{11}, 124 (2015)}.

\bibitem{Huse-review}
R. Nandkishore and D. A. Huse,
\href{https://www.annualreviews.org/doi/abs/10.1146/annurev-conmatphys-031214-014726}
{Annu. Rev. Condens. Matter Phys. \textbf{6}, 15 (2015)}.

\bibitem{Schmiedmayer}
T. Langen, R. Geiger, and J. Schmiedmayer,
\href{https://www.annualreviews.org/doi/abs/10.1146/annurev-conmatphys-031214-014548}
{Annu. Rev. Condens. Matter Phys. \textbf{6}, 201 (2015)}.

\bibitem{Grover-PRX}
J. R. Garrison and T. Grover,
\href{https://journals.aps.org/prx/abstract/10.1103/PhysRevX.8.021026}
{Phys. Rev. X \textbf{8}, 021026 (2018)}.

\bibitem{Deutsch2013}
J. M. Deutsch, H. Li, and A. Sharma,
\href{https://journals.aps.org/pre/abstract/10.1103/PhysRevE.87.042135}
{Phys. Rev. E \textbf{87}, 042135 (2013)}.

\bibitem{Goold}
M. T. Mitchison, A. Purkayastha, M. Brenes, A. Silva, and J. Goold,
\href{https://arxiv.org/abs/2103.16601}
{arXiv:2103.16601}

\bibitem{Haque-Temp}
P. C. Burke, G. Nakerst, and M. Haque,
\href{https://arxiv.org/abs/2111.05083}
{arXiv:2111.05083}.

\bibitem{vN}
J. von Neumann, \textit{Mathematische Grundlagen der Quantenmechanik} (Springer, Berlin, 1932); \textit{Mathematical Foundations of Quantum Mechanics} (Princeton University Press, Princeton, 1955).

\bibitem{Yang2015}
H.-H. Lai and K. Yang,
\href{https://journals.aps.org/prb/abstract/10.1103/PhysRevB.91.081110}
{Phys. Rev. B \textbf{91}, 081110(R) (2015)}.

\bibitem{Yang2018}
C. Tian, K. Yang, P. Fang, H.-J. Zhou, and J. Wang,
\href{https://journals.aps.org/pre/abstract/10.1103/PhysRevE.98.060103}
{Phys. Rev. E \textbf{98}, 060103(R) (2018)}.

\bibitem{TianYang}
C. Tian and K. Yang,
\href{https://journals.aps.org/prb/abstract/10.1103/PhysRevB.104.174302}
{Phys. Rev. B {\bf 104}, 174302 (2021).}

\bibitem{Horodecki-RMP}
R. Horodecki, P. Horodecki, M. Horodecki, and K. Horodecki,
\href{https://journals.aps.org/rmp/abstract/10.1103/RevModPhys.81.865}
{Rev. Mod. Phys. \textbf{81}, 865 (2009)}.

\bibitem{Deutsch2010}
J. M. Deutsch,
\href{https://iopscience.iop.org/article/10.1088/1367-2630/12/7/075021}
{New J. Phys. \textbf{12}, 075021 (2010)}.

\bibitem{Baez}
J. C. Baez and M. Stay,
\href{https://doi.org/10.1017/S0960129511000521}
{Math. Struct. Comput. Sci., \textbf{22}, 771 (2012)}.

\bibitem{Deutsch-en}
D. \v{S}afr\'{a}nek, A. Aguirre, J. Schindler, and J. M. Deutsch,
\href{https://link.springer.com/article/10.1007%2Fs10701-021-00498-x}
{Found. Phys. \textbf{51}, 101 (2021)}, and references therein.

\bibitem{Winter-PRXQ}
P. Strasberg and A. Winter,
\href{https://journals.aps.org/prxquantum/abstract/10.1103/PRXQuantum.2.030202}
{PRX Quantum \textbf{2}, 030202 (2021)}.

\bibitem{Holecek}
M. Hole\v{c}ek,
\href{https://arxiv.org/abs/2103.16913}
{arXiv:2103.16913}

\bibitem{Zurek1989}
W. H. Zurek,
\href{https://journals.aps.org/pra/abstract/10.1103/PhysRevA.40.4731}
{Phys. Rev. A \textbf{40}, 4731 (1989)}.

\bibitem{Zurek-nature}
W. H. Zurek,
\href{https://www.nature.com/articles/341119a0}
{Nature \textbf{341}, 119 (1989)}.

\bibitem{Zurek-book}
\textit{Complexity, Entropy And The Physics Of Information,} 1st ed., edited by W. H. Zurek (CRC, Boca Raton, FL 1990).

\bibitem{Caves}
C. M. Caves,
\href{https://journals.aps.org/pre/abstract/10.1103/PhysRevE.47.4010}
{Phys. Rev. E \textbf{47}, 4010 (1993)}.

\bibitem{Susskind1}
A. R. Brown, L. Susskind, and Y. Zhao,
\href{https://journals.aps.org/prd/abstract/10.1103/PhysRevD.95.045010}
{Phys. Rev. D \textbf{95}, 045010 (2017)}.

\bibitem{Susskind2}
A. R. Brown and L. Susskind,
\href{https://journals.aps.org/prd/abstract/10.1103/PhysRevD.97.086015}
{Phys. Rev. D \textbf{97}, 086015 (2018)}.


\bibitem{Svozil}
K. Svozil, J Univers. Comput Sci. \textbf{2}, 311 (1996).

\bibitem{Vitanyi2000}
P. Vit\'{a}nyi, in 
\textit{Proceedings 15th Annual IEEE Conference on Computational Complexity,}
\href{https://ieeexplore.ieee.org/document/856757}
{(IEEE Computer Society, Los Alamitos, CA, 2000), pp.263-270.}


\bibitem{Vitanyi2001}
P. M. B. Vit\'{a}yni,
\href{https://ieeexplore.ieee.org/document/945258}
{IEEE Trans. Inf. Theory \textbf{47}, 2464 (2001)}.

\bibitem{Gacs}
P. G\'{a}cs,
\href{https://iopscience.iop.org/article/10.1088/0305-4470/34/35/312}
{J. Phys. A: Math. Gen. \textbf{34}, 6859 (2001)}.

\bibitem{Berthiaume}
A. Berthiaume, W. van Dam, and S. Laplante,
\href{https://www.sciencedirect.com/science/article/pii/S0022000001917659}
{J. Comput. Syst. Sci. \textbf{63}, 201 (2001)}.

\bibitem{Yamakami}
T. Yamakami, Proc. 14th ISAAC. Springer's LNCS, Vol. 2906, pp.117, (2003),
\href{https://arxiv.org/abs/quant-ph/0308072}
{arXiv:quant-ph/0308072}.

\bibitem{Benatti}
F. Benatti, T. Kr\"{u}ger, M. M\"{u}ller, R. Siegmund-Schultze, and A. Szko\l a,
\href{https://link.springer.com/article/10.1007/s00220-006-0027-z}
{Commun. Math. Phys. \textbf{265}, 437 (2006)}.

\bibitem{Benatti2}
F. Benatti, 
\href{https://doi.org/10.1007/s11047-006-9017-5}
{Nat. Comput. \textbf{6}, 133 (2007)}.

\bibitem{Mueller-thesis}
M. M\"{u}ller, Ph.D. thesis, Technical University of Berlin, 2007,
\href{https://arxiv.org/abs/0712.4377}
{arXiv:0712.4377}.

\bibitem{Rogers}
C. Rogers, V. Vedral and R. Nagarajan,
\href{https://www.worldscientific.com/doi/10.1142/S021974990800375X}
{Int. J. Quantum Inf. \textbf{06}, 907 (2008)}.

\bibitem{Mora-PRL}
C. E. Mora and H. J. Briegel,
\href{https://journals.aps.org/prl/abstract/10.1103/PhysRevLett.95.200503}
{Phys. Rev. Lett. \textbf{95}, 200503 (2005)}.

\bibitem{Mora2006}
C. E. Mora and H. J. Briegel,
\href{https://www.worldscientific.com/doi/10.1142/S0219749906002043}
{Int. J. Quantum Inf. \textbf{04}, 715 (2006)}.

\bibitem{Mora2007}
C. E. Mora, H. J. Briegel, and B. Kraus,
\href{https://www.worldscientific.com/doi/abs/10.1142/S0219749907003171}
{Int. J. Quantum Inf. \textbf{05}, 729 (2007)}.

\bibitem{Eisert2016}
Y. Ge and J. Eisert,
\href{https://iopscience.iop.org/article/10.1088/1367-2630/18/8/083026}
{New J. Phys. \textbf{18}, 083026 (2016)}.

\bibitem{Aguero}
M. G. Kovalsky, A. A. Hnilo, and M. B. Ag\"{u}ero,
\href{https://journals.aps.org/pra/abstract/10.1103/PhysRevA.98.042131}
{Phys. Rev. A \textbf{98}, 042131 (2018)}.

\bibitem{Eisert2021}
J. Eisert,
\href{https://journals.aps.org/prl/abstract/10.1103/PhysRevLett.127.020501}
{Phys. Rev. Lett. \textbf{127}, 020501 (2021)}.

\bibitem{Hnilo}
M. Nonaka, M. Ag\"{u}ero, M. Kovalsky, and A. Hnilo,
\href{https://arxiv.org/abs/1908.10794}
{arXiv:1908.10794}

\bibitem{Bhojraj}
T. Bhojraj,
\href{https://www.sciencedirect.com/science/article/pii/S0304397521002814}
{Theor. Comput. Sci. \textbf{875}, 65 (2021)}.

\bibitem{Kazemi}
J. Kazemi and H. Weimer,
\href{https://arxiv.org/abs/2106.07673}
{arXiv:2106.07673}

\bibitem{Kaltchenko}
A. Kaltchenko,
\href{https://arxiv.org/abs/2110.05937}
{arXiv:2110.05937}.

\bibitem{Mueller-QKC-KC}
For a classical string, its quantum Kolmogorov complexity and classical Kolmogorov complexity differ by an $O(1)$ term. See M. M\"{u}ller,
\href{https://www.worldscientific.com/doi/10.1142/S0219749909005456}
{Int. J. Quant. Inf. \textbf{7}, 701-711 (2009)}.

\bibitem{Serbyn-review}
For a review, see M. Serbyn, D. A. Abanin, and Z. Papi\'{c},
\href{https://www.nature.com/articles/s41567-021-01230-2}
{Nat. Phy. \textbf{17}, 675 (2021)} and the references therein.

\bibitem{Solomonoff1964}
R. J. Solomonoff,
\href{https://www.sciencedirect.com/science/article/pii/S0019995864902232}
{Inf. Control \textbf{7}, 1 (1964)};
\href{https://www.sciencedirect.com/science/article/pii/S0019995864901317}
{Inf. Control \textbf{7}, 224 (1964)}.

\bibitem{Kolmogorov1965}
A. N. Kolmogorov, Probl. Inf. Transm. \textbf{1}, 3 (1965).

\bibitem{Chatin1987}
G. J. Chaitin, \textit{Information, Randomness \emph{\&} Incompleteness} (World Scientific, Singapore, 1987).

\bibitem{Li-Vitanyi}
M. Li and P. Vit\'{a}nyi, \textit{An Introduction to Kolmogorov Complexity and Its Applications}, 4th ed. (Springer, Berlin, 2019).

\bibitem{footnote-K}
To be more precise, one should consider the prefix Kolmogorov complexity, $K(x)$. In a self-delimiting program, $x$ is encoded as a prefix-free code~\cite{Li-Vitanyi}. Using $K(x)$, the joint Kolmogorov complexity $K(x_1,x_2)$ satisfies the subadditive condition: $K(x_1,x_2)\leq K(x_1)+K(x_2)+O(1)$. This feature provides a more rigorous connection between the Kolmogorov complexity and Shannon entropy~\cite{Shannon}. Since the self-delimiting program $p^*$ requires an additional $O(\log{[l(p)]})$ bits to store the length of $p$ where $p$ is the non self-delimiting program, $K(x)$ and $C(x)$ are different. Compared with the leading order term $l(p)$, $|K(x)-C(x)|$ is always a subleading-order term that does not affect our results in the main text. 

\bibitem{footnote-log}
To follow the convention in information theory, $\log$ is defined as the logarithm with base two.

\bibitem{pigeonhole}
P.G.L. Dirichlet and R. Dedekind, \textit{Lectures on Number Theory} (American
Mathematical Society, Providence, RI, 1999).

\bibitem{Shannon}
C. E. Shannon,
\href{https://ieeexplore.ieee.org/document/6773024}
{Bell Syst. Tech. J. \textbf{27}, 379 (1948)}.

\bibitem{Hastings}
M. B. Hastings,
\href{https://iopscience.iop.org/article/10.1088/1742-5468/2007/08/P08024}
{J. Stat. Mech.: Theory Exp. \textbf{2007}, P08024 (2007)}.

\bibitem{Amico-RMP}
L. Amico, R. Fazio, A. Osterloh, and V. Vedral,
\href{https://journals.aps.org/rmp/abstract/10.1103/RevModPhys.80.517}
{Rev. Mod. Phys. \textbf{80}, 517 (2008)}.

\bibitem{Eisert-RMP}
J. Eisert, M. Cramer, and M. B. Plenio,
\href{https://journals.aps.org/rmp/abstract/10.1103/RevModPhys.82.277}
{Rev. Mod. Phys. \textbf{82}, 277 (2010)}.

\bibitem{Wolf2006}
M. M. Wolf,
\href{https://journals.aps.org/prl/abstract/10.1103/PhysRevLett.96.010404}
{Phys. Rev. Lett. \textbf{96}, 010404 (2006)}.

\bibitem{GK2006}
D. Gioev and I. Klich, \href{https://journals.aps.org/prl/abstract/10.1103/PhysRevLett.96.100503}
{Phys. Rev. Lett. \textbf{96}, 100503 (2006)}.

\bibitem{Swingle2010}
B. Swingle,
\href{https://journals.aps.org/prl/abstract/10.1103/PhysRevLett.105.050502}
{Phys. Rev. Lett. \textbf{105}, 050502 (2010)}.

\bibitem{Ding-PRX}
W. Ding, A. Seidel, and K. Yang,
\href{https://journals.aps.org/prx/abstract/10.1103/PhysRevX.2.011012}
{Phys. Rev. X \textbf{2}, 011012 (2012)}.

\bibitem{footnote-violation}
Note that a violation of area law also occurs in one-dimensional critical systems~\cite{Cardy, Kitaev2003}.

\bibitem{footnote-many}
When $m\sim n$, the system should be viewed as being in a highly excited state. The Kolmogorov complexity of the corresponding occupation pattern should scale as $N$.

\bibitem{Diestel-graph}
R. Diestel, \textit{Graph Theory}, 5th ed. (Springer, Berlin, 2017).

\bibitem{K-graph}
A. Farzaneh, J. P. Coon, and M.-A. Badiu,
\href{https://www.mdpi.com/1099-4300/23/12/1604}
{Entropy \textbf{23}, 1604 (2021)}.

\bibitem{Potter}
A. C. Potter,
\href{https://arxiv.org/abs/1408.1094}
{arXiv:1408.1094}.

\bibitem{Pouranvari}
M. Pouranvari, Y. Zhang, and K. Yang,
\href{https://www.hindawi.com/journals/acmp/2015/397630/}
{Adv. Condensed Matter Phys. \textbf{2015}, 397630 (2015)}.

\bibitem{footnote-generic}
For a generic translationally invariant Hamiltonian with energy dispersion $\epsilon(\mathbf{k})$, the momentum of the ground state need not be at $\mathbf{k}=\mathbf{0}$.

\bibitem{Klich2006}
I. Klich, G. Refael, and A. Silva,
\href{https://journals.aps.org/pra/abstract/10.1103/PhysRevA.74.032306}
{Phys. Rev. A \textbf{74}, 032306 (2006)}.

\bibitem{Ding2009}
W. Ding and K. Yang,
\href{https://journals.aps.org/pra/abstract/10.1103/PhysRevA.80.012329}
{Phys. Rev. A \textbf{80}, 012329 (2009)}.

\bibitem{Ding2008}
W. Ding, N. E. Bonesteel, and K. Yang,
\href{https://journals.aps.org/pra/abstract/10.1103/PhysRevA.77.052109}
{Phys. Rev. A \textbf{77}, 052109 (2008)}.

\bibitem{Grover}
M. A. Metlitski and T. Grover,
\href{https://arxiv.org/abs/1112.5166}
{arXiv:1112.5166}.

\bibitem{Wen-TEE}
M. Levin and X.-G. Wen,
\href{https://journals.aps.org/prl/abstract/10.1103/PhysRevLett.96.110405}
{Phys. Rev. Lett. \textbf{96}, 110405 (2006)}.

\bibitem{Kitaev-TEE}
A. Kitaev and J. Preskill,
\href{https://journals.aps.org/prl/abstract/10.1103/PhysRevLett.96.110404}
{Phys. Rev. Lett. \textbf{96}, 110404 (2006)}.

\bibitem{Grunwald}
P. Grunwald and P. Vitanyi,
\href{https://arxiv.org/abs/cs/0410002}
{arXiv:cs/0410002}.

\bibitem{strong-ETH}
G. Biroli, C. Kollath, and A. M. L\"{a}uchli,
\href{https://journals.aps.org/prl/abstract/10.1103/PhysRevLett.105.250401}
{Phys. Rev. Lett. \textbf{105}, 250401 (2010)}.

\bibitem{Kitaev2003}
G. Vidal, J. I. Latorre, E. Rico, and A. Kitaev, 
\href{https://journals.aps.org/prl/abstract/10.1103/PhysRevLett.90.227902}
{Phys. Rev. Lett. \textbf{90}, 227902 (2003)}.

\bibitem{Cardy}
P. Calabrese and J. Cardy,
\href{https://iopscience.iop.org/article/10.1088/1742-5468/2004/06/P06002}
{J. Stat. Mech.: Theory Exp. \textbf{2004}, P06002 (2004)}.


\end{thebibliography}
\end{document}